\documentclass[aps,11pt,preprintnumbers,nofootinbib,floatfix]{revtex4}

\usepackage{graphicx}
\usepackage{epsfig}
\usepackage{dcolumn}
\usepackage{subfigure}
\usepackage{multirow}

\def\be{\begin{equation}}
\def\ee{\end{equation}}
\def\bea{\begin{eqnarray}}
\def\eea{\end{eqnarray}}

\def\no{\nonumber}

\newcommand{\omits}[1]{}

\begin{document}
\title{Entropy of higher-dimensional charged de Sitter black holes and phase transition}

\author{Ren Zhao\footnote{Email: zhao2969@sina.com} and Li-Chun Zhang\footnote{Email: sub\_ 2016@163.com}}
\medskip

\affiliation{Institute of Theoretical Physics, Shanxi Datong
University, Datong 037009, China\\
Department of Physics, Shanxi Datong
University, Datong 037009, China}

\begin{abstract}

From a new perspective, we discuss the thermodynamic entropy of $n+2$-dimensional Reissner-Nordstr\"{o}m-de Sitter(RNdS) black hole and analyze the phase transition of the effective thermodynamic system.
Considering the correlations between the black hole event horizon and the cosmological horizon, we conjecture that the total entropy of the RNdS black hole should contain an extra term  besides the sum of the entropies of the two horizons. In the lukewarm case, the effective temperature of the RNdS black hole is the same as that of the black hole horizon and the cosmological horizon. Under this condition, we obtain the extra contribution to the total entropy. With the corrected entropy, we derive other effective thermodynamic quantities and analyze the phase transition of the RNdS black hole in analogy to the usual thermodynamic system.

\end{abstract}

\maketitle

\section{Introduction}

Black holes are exotic objects in the theory of classical and quantum
gravity. Even more surprising is their connection with the laws of standard
thermodynamics. Since black hole thermodynamics is expected to play a role
in any meaningful theory of gravity, therefore it will be a natural question
to ask whether the thermodynamic properties of black holes are modified if
higher dimensional corrections are incorporated in the Einstein-Hilbert
action. One can expect a similar situation to appear in an effective theory
of quantum gravity, such as string theory.

Black holes in different various dimensional sapcetime with different
geometric properties have been drawing many interests. Many physical
properties of black holes are related to its thermodynamic properties, such
as entropy, Hawking radiation. Recently, the idea of including the variation
of the cosmological constant $\Lambda $ in the first law of black hole
thermodynamics has attained increasing attention \cite{Kastor-2009,Kubiznak:2012,Dolan:2011,Gunasekaran,Banerjee:2011raa,Cai:2013,Zhao:2013,Ma.2014,
Dolan.2014,Ma.2014b,Ma.2015,Ma.2015b,ZhaoHH.2015,BEP:2015,Dehghani,YXL:2016,Hendi:2016,Panahiyan:2016,Chabab:2016,Miao:2016,Sadeghi:2016,Meng:2016,Poshteh:2017,BEP:2017,Ma:2017}. Comparing the
thermodynamic quantities in AdS black holes with those of conventional
thermodynamic system, the P-V criticalities of these black holes have been
extensively studied. It is shown that the phase structure, critical exponent
and Clapeyron equation of the AdS black holes are similar to those of a van
der Waals liquid/gas system.

As is well known, de Sitter black holes can have both the black hole event
horizon and the cosmological horizon. Both the horizons can radiate, however
their temperatures are different generally. Therefore, the whole de Sitter
black hole system is thermodynamically unstable. We also know that the two
horizons both satisfy the first law of thermodynamics and the corresponding
entropies both satisfy the area law\cite{Cai:2002,Sekiwa,Urano}. In recent years, the studies on
the thermodynamic properties of de Sitter space have aroused wide
attention\cite{Cai:2002,Sekiwa,Urano,Gombero:2003,Mann:2005,Padmanabhan:2007,Bhattacharya,Katsuragawa,MAA1,MAA2,Hajian,Kubiznak,McInerney,Li-2016}. In the early inflation epoch, the universe is a quasi-de
Sitter spacetime. If the cosmological constant is the dark energy, our
universe will evolve to a new de Sitter phase.

Because the two horizons are expressed by the same parameters: the mass M,
electric charge Q and the cosmological constant $\Lambda $, they should be
dependent each other. Taking into account of the correlations between the two horizons
is very important for the description of the thermodynamic properties of de
Sitter black holes. Previous works, such as Refs.\cite{Mellor,Romans,Cai:1998,BPD:2013,ZR:2014,ZLC:2014,ZHH,MMS:2014,Brenna,Poshteh2,Ma,Guo2,ZLC,Guo1,MYQ}, considered that the
entropy of the de Sitter black holes is the sum of the black hole entropy
and the entropy of the cosmological horizon. Based on this consideration,
the effective thermodynamic quantities and phase transition are analyzed. It
shows that de Sitter black holes have the similar critical behaviors to
those of black holes in anti-de Sitter space. However, considering the
correlation or entanglement between the event horizon and the cosmological
horizon, the total entropy of the charged black hole in de Sitter space is
no longer simply $S=S_+ +S_c $, but should include an extra term from the
contribution of the correlations of the two horizons\cite{ZLC:2016,LHF:2017}.

In this paper, we study the $(n+2)$-dimensional Reissner-Nordstr\"{o}m-dS black
hole by considering the correlation of the black hole horizon and the
cosmological horizon. In the Sec.II, we review the various thermodynamic
quantities on the both horizons and give the condition under which the
temperatures of the two horizons are equal. In Sec. III, we derive the
effective thermodynamic quantities and propose the expression of the whole
entropy. In Sec. IV, the phase transition of the higher-dimensional RN-dS
black hole is studied according to the Ehrenfest's equations. At last, we
will give the conclusions. (we use the units $\hbar =k_B =c=1)$.

\section{Lukewarm $(n+2)$-dimensional Reissner-Nordstrom solutions in de Sitter
space}

The line element of the $(n+2)$-dimensional RNdS black hole is given by\cite{Cai:2002}
\begin{equation}
\label{eq1}
ds^2=-f(r)dt^2+f^{-1}(r)dr^2+r^2d\Omega _n^2 ,
\end{equation}
where
\begin{equation}
\label{eq2}
f(r)=1-\frac{\omega _n M}{r^{n-1}}+\frac{n\omega _n^2
Q^2}{8(n-1)r^{2n-2}}-\frac{r^2}{l^2},
\quad
\omega _n =\frac{16\pi G}{nVol(S^n)}.
\end{equation}
Here $G$ is the gravitational constant in $(n+2)$ dimensions, $l$ is the
curvature radius of dS space, $Vol(S^n)$ denotes the volume of a unit
$n$-sphere $d\Omega _n^2 $, $M$ is an integration constant and $Q$ is the
electric/magnetic charge of Maxwell field. For general $M$ and $Q$, the
equation $f(r)=0$ may have four real roots. Three of them are real: the
largest one is the cosmological horizon $r_c $, the smallest is the inner
(Cauchy) horizon of black hole, the middle one is the event horizon $r_+ $
of black hole. Some thermodynamic quantities associated with the
cosmological horizon are
\bea\label{eq3}
T_c &=&\frac{1}{4\pi r_c }\left[ {-(n-1)+(n+1)\frac{r_c^2 }{l^2}+\frac{n\omega
_n^2 Q^2}{8r_c^{2n-2} }} \right],\\
S_c &=&\frac{r_c^n Vol(S^n)}{4G},\\
\Phi _c &=&\frac{n}{4(n-1)}\frac{\omega _n Q}{r_c^{n-1} },
\eea
where $\Phi _c $ is the chemical potential conjugate to the charge $Q$. The
first law of thermodynamics of the cosmological horizon is \cite{BPD:2013}
\begin{equation}
\label{eq4}
dM=T_c dS_c +V_c dP+\Phi _c dQ,
\end{equation}
\begin{equation}
\label{eq5}
P=-\frac{n(n+1)}{16\pi l^2},
\quad
V_c =\frac{Vol(S^n)}{(n+1)}r_c^{n+1} .
\end{equation}
For the black hole horizon, associated thermodynamic quantities are
\bea\label{eq6}
T_+ &=&\frac{1}{4\pi r_+ }\left[ {(n-1)-(n+1)\frac{r_+^2 }{l^2}-\frac{n\omega
_n^2 Q^2}{8r_+^{2n-2} }} \right],\\
S_+ &=&\frac{r_+^n Vol(S^n)}{4G}, \\
\Phi _+ &=&\frac{n}{4(n-1)}\frac{\omega _n Q}{r_+^{n-1} }.
\eea
The first law of thermodynamics of the black hole horizon is\cite{BPD:2013}
\begin{equation}
\label{eq7}
dM=T_+ dS_+ +V_+ dP+\Phi _+ dQ,
\end{equation}
\begin{equation}
\label{eq8}
V_+ =\frac{Vol(S^n)}{(n+1)}r_+^{n+1} .
\end{equation}
In the following, we find the ``lukewarm'' $(n+2)$-dimensional RN solutions
which realize this state of affairs, that is, describing an outer black hole
horizon at radius $r_+ $ and a de Sitter edge at radius $r_c $, with the
same Hawking temperature at $r_+ $ and $r_c $. In terms of the metric
function $f(r)$, the algebraic problem is \cite{Cai:1998,Romans,Mellor}
\begin{equation}
\label{eq9}
f(r_c )=f(r_+ )=0;
\quad
f'(r_c )=\pm f'(r_+ ),
\end{equation}
where the minus sign is appropriate, since there should be no roots of
$f(r)$ between $r_+ $ and $r_c $.

According to $f(r_+ )=f(r_c )=0$, one can derive
\begin{equation}
\label{eq10}
\frac{1}{l^2}=\frac{r_c^{n-1} -r_+^{n-1} }{r_c^{n+1}
-r_+^{n+1} }\left[ {1-\frac{A}{(n-1)}} \right],
\end{equation}
and
\begin{equation}
\label{eq11}
\omega _n M=r_c^{n-1} x^{n-1}\frac{1-x^2}{1-x^{n+1}}+\frac{n\omega _n^2
Q^2}{8(n-1)r_c^{n-1} x^{n-1}}\frac{1-x^{2n}}{1-x^{n+1}}.
\end{equation}
where we have set $\frac{n\omega _n^2 Q^2}{8(r_c r_+ )^{n-1}}=\frac{n\omega _n^2
Q^2}{8r_c^{2n-2} x^{n-1}}=A$ and $x=r_+ /r_c $.

From $T_+ =T_c $, we can get
\begin{equation}
\label{eq12}
\frac{r_c^{n-1} -r_+^{n-1} }{r_c^{n+1} -r_+^{n+1} }-\frac{(n-1)}{(n+1)r_c
r_+ }=\frac{n\omega _n^2 Q^2}{8}F(r_+ ,r_c ,n).
\end{equation}
where
\begin{equation}
\label{eq13}
F(r_+ ,r_c ,n)=\frac{(r_c^{n-1} -r_+^{n-1} )}{(n-1)r_c^{n-1} r_+^{n-1}
(r_c^{n+1} -r_+^{n+1} )}-\frac{r_c^{2n-1} +r_+^{2n-1} }{(n+1)r_+^{2n-1}
r_c^{2n-1} (r_c +r_+ )}.
\end{equation}
Substituting Eqs. (\ref{eq10}) and (\ref{eq11}) into Eqs. (\ref{eq3}) and (\ref{eq6}), the lukewarm
temperature $T_{c+}$ is
\bea\label{eq14}
T_{c+}&=&\frac{T_+ +T_c }{2}\no \\
&=&\frac{(n-1)(1-x)}{4\pi r_c x}-\frac{(1-x^{2n})}{4\pi r_c
x^{2n-1}(1+x)r_c^{2n} F(r_c ,x,n)}\left[
{\frac{1-x^{n-1}}{1-x^{n+1}}-\frac{(n-1)}{(n+1)x}} \right],
\eea
where
\begin{equation}
\label{eq15}
F(r_c ,x,n)=\frac{(1-x^{n-1})}{(n-1)r_c^{£n}
x^{n-1}(1-x^{n+1})}-\frac{1+x^{2n-1}}{(n+1)r_c^{2n} x^{2n-1}(1+x)}.
\end{equation}

When the cosmological constant satisfies Eq.(\ref{eq10}), and the electric charge
Q satisfies Eq.(\ref{eq12}), the temperatures of the two horizons are equal, which
is given in Eq.(\ref{eq14}).

\begin{figure}[!htbp]
\includegraphics[width=7cm,keepaspectratio]{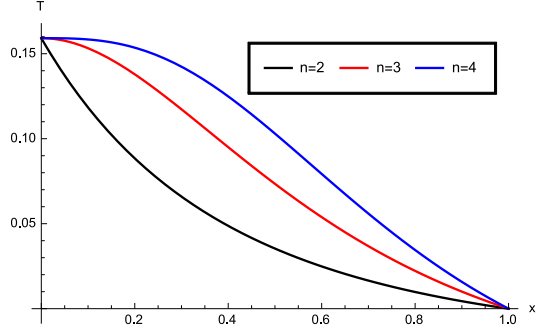}
\caption{ The temperature of lukewarm black hole as function of $x$ for different spacetime dimensions. We have set $r_c=1$.}\label{figTx}
\end{figure}

As is depicted in Fig.\ref{figTx}, in the lukewarm case, the temperature of the horizons increases with the dimension of spacetime and monotonically decreases with the increase of $x$. This means that the closer the two horizons are, the lower of their temperature will be.

\section{ Entropy of the $(n+2)$-dimensional RNdS black hole}

The thermodynamic quantities of $(n+2)$-dimensional RNdS black hole satisfy\cite{ZR:2014,ZLC:2014}
\begin{equation}
\label{eq15}
dM=T_{eff} dS-P_{eff} dV+\Phi _{eff} dQ,
\end{equation}
where the thermodynamic volume is \cite{McInerney,BPD:2013,Brenna}
\be\label{Volume}
V=V_c -V_+ =\frac{Vol(S^n)}{(n+1)}(r_c^{n+1} -r_b^{n+1}
)=\frac{Vol(S^n)}{(n+1)}r_c^{n+1} (1-x^{n+1}).
\ee

The effective temperature, the effective pressure and the effective electric
potential are respectively
\begin{equation}
\label{eq16}
T_{eff} =\left( {\frac{\partial M}{\partial S}} \right)_{Q,V} =\frac{\left(
{\frac{\partial M}{\partial x}} \right)_{r_c } \left( {\frac{\partial
V}{\partial r_c }} \right)_x -\left( {\frac{\partial V}{\partial x}}
\right)_{r_c } \left( {\frac{\partial M}{\partial r_c }} \right)_x }{\left(
{\frac{\partial S}{\partial x}} \right)_{r_c } \left( {\frac{\partial
V}{\partial r_c }} \right)_x -\left( {\frac{\partial V}{\partial x}}
\right)_{r_c } \left( {\frac{\partial S}{\partial r_c }} \right)_x },
\end{equation}
\be
P_{eff} =-\left( {\frac{\partial M}{\partial V}} \right)_{Q,S}
=-\frac{\left( {\frac{\partial M}{\partial x}} \right)_{r_c } \left(
{\frac{\partial S}{\partial r_c }} \right)_x -\left( {\frac{\partial
S}{\partial x}} \right)_{r_c } \left( {\frac{\partial M}{\partial r_c }}
\right)_x }{\left( {\frac{\partial V}{\partial x}} \right)_{r_c } \left(
{\frac{\partial S}{\partial r_c }} \right)_x -\left( {\frac{\partial
S}{\partial x}} \right)_{r_c } \left( {\frac{\partial V}{\partial r_c }}
\right)_x },
\ee
\be
\Phi _{eff} =\left( {\frac{\partial M}{\partial Q}}
\right)_{S,V}.
\ee

For a system composed of two subsystems, the total entropy should be the simple sum of the entropies of the two subsystems if there is no interactions between them. When correlation exists between the two subsystems, the total entropy should contain an extra contribution coming from the correlations between the two subsystems. Considering the correlation between the two horizons, we conjecture that the
entropy of the $(n+2)$-dimensional RNdS black hole should take the form of
\be\label{CS}
S=\frac{Vol(S^n)}{4G}r_c^n (1+x^n+f(x)),
\ee
where the undefined function $f(x)$ represents the extra contribution from the
correlations of the two horizons. Next we try to determine the concrete form of
$f(x)$.

Substituting Eqs. (\ref{eq11}), (\ref{Volume}) and (\ref{CS}) into Eq. (\ref{eq16}), we can get
\begin{equation}
\label{Teff}
T_{eff} =\frac{4G}{Vol(S^n)r_c^{n-1} }\frac{\frac{1}{r_c }\left(
{\frac{\partial M}{\partial x}} \right)_{r_c } \left( {1-x^{n+1}}
\right)+\left( {\frac{\partial M}{\partial r_c }} \right)_x
x^n}{(nx^{n-1}+f'(x))(1-x^{n+1})+nx^n(1+x^n+f(x))}.
\end{equation}
From Eq.(\ref{eq11}), we can derive
\be
\left( {\frac{\partial M}{\partial r_c }} \right)_{x}=(n-1)\frac{r_c
^{n-2}x^{n-1}}{\omega _n }\left( {\frac{1-x^2}{1-x^{n+1}}}
\right)£\frac{n\omega _n Q^2(1-x^{2n})}{8r_c^n x^{n-1}(1-x^{n+1})},
\ee
and
\be
\left( {\frac{\partial M}{\partial x}} \right)_{r_c } =\frac{r_c
^{n-1}}{\omega _n
}\frac{(n-1)x^{n-2}-(n+1)x^n+2x^{2n-1}}{(1-x^{n+1})^2}+\frac{n\omega _n
Q^2(2nx^{n+1}-(n-1)-(n+1)x^{2n})}{8(n-1)r_c^{n-1} x^n(1-x^{n+1})^2}.
\ee

When the temperatures of the two horizons are the same, the charge $Q$ satisfies Eq.(\ref{eq12}). Thus, we can derive the effective temperature $T_{eff}$ in the lukewarm case:
\be\label{TeffLM}
T_{eff} =\frac{n}{4\pi r_c}\frac{A+B}{(nx^{n-1}+f'(x))(1-x^{n+1})+nx^n(1+x^n+f(x))},
\ee
where
\[
A=\frac{(n-1)x^{n-2}-(n+1)x^n+2x^{2n-1}}{(1-x^{n+1})}+\frac{2nx^{n+1}-(n-1)-(n+1)x^{2n}}{(n-1)x^n(1-x^{n+1})F_0 (r_c ,x,n)}\left(
{\frac{(1-x^{n-1})}{££-x^{n+1})}-\frac{(n-1)}{(n+1)x}} \right),
\]
\[
B=(n-1)x^{2n-1}\left( {\frac{1-x^2}{1-x^{n+1}}}
\right)£\frac{x(1-x^{2n})}{(1-x^{n+1})F_0 (r_c ,x,n)}\left(
{\frac{(1-x^{n-1})}{££-x^{n+1})}-\frac{(n-1)}{(n+1)x}} \right),
\]
with
\[
F_0 (r_c
,x,n)=\frac{(1-x^{n-1})}{(n-1)x^{n-1}(1-x^{n+1})}-\frac{1+x^{2n-1}}{(n+1)x^{2n-1}(1+x)}.
\]

When the two horizons have the same temperature, we think the effective
temperature of the system should have the same value, namely
\begin{equation}
\label{eq18}
T_{eff} =T_c =T_+ =T_{c+}.
\end{equation}
According to Eqs. (\ref{eq14}) and (\ref{TeffLM}), we derive the equations about $f(x)$:
\be\label{fxeq}
f'(x)-\frac{n x^nf(x)}{x^{n+1}-1}=\frac{n x^n \left(2 x^{n+1}+x^n-1\right)}{\left(x^{n+1}-1\right)^2}.
\ee
For $n=2,~n=3,~n=4$, the field equations about $f(x)$ are respectively :
\bea\label{eqn2}
f'(x)+\frac{2 x^2 f(x)}{1-x^3}&=&\frac{2 x^2 \left(2 x^3+x^2-1\right)}{\left(1-x^3\right)^2},\\
f'(x)+\frac{3 x^3 f(x)}{1-x^4}&=&\frac{3 x^3 \left(2 x^4+x^3-1\right)}{\left(1-x^4\right)^2},\\
f'(x)+\frac{4 x^4 f(x)}{1-x^5}&=&\frac{4 x^4 \left(2 x^5+x^4-1\right)}{\left(1-x^5\right)^2}.
\eea
And the solutions for these equations are respectively:
\bea
f(x)&=&\frac{8}{5} \left(1-x^3\right)^{2/3}-\frac{2 \left(x^5+5 x^3-4\right)}{5 \left(x^3-1\right)},\\
f(x)&=&\frac{11}{7} \left(1-x^4\right)^{3/4}+\frac{-3 x^6+3 x^5-3 x^4-11 x^3+11 x^2-11 x+11}{7 \left(x^3-x^2+x-1\right)}, \\
f(x)&=&\frac{14}{9} \left(1-x^5\right)^{4/5}+\frac{2 \left(-2 x^9-9 x^5+7\right)}{9 \left(x^5-1\right)},
\eea
where we have taken the boundary condition $f(0)=0$, because $x=0$ means the absence of the black hole horizon and thus no correlation between the black hole horizon and the cosmological horizon.

\begin{figure}[!htbp]
\includegraphics[width=7cm,keepaspectratio]{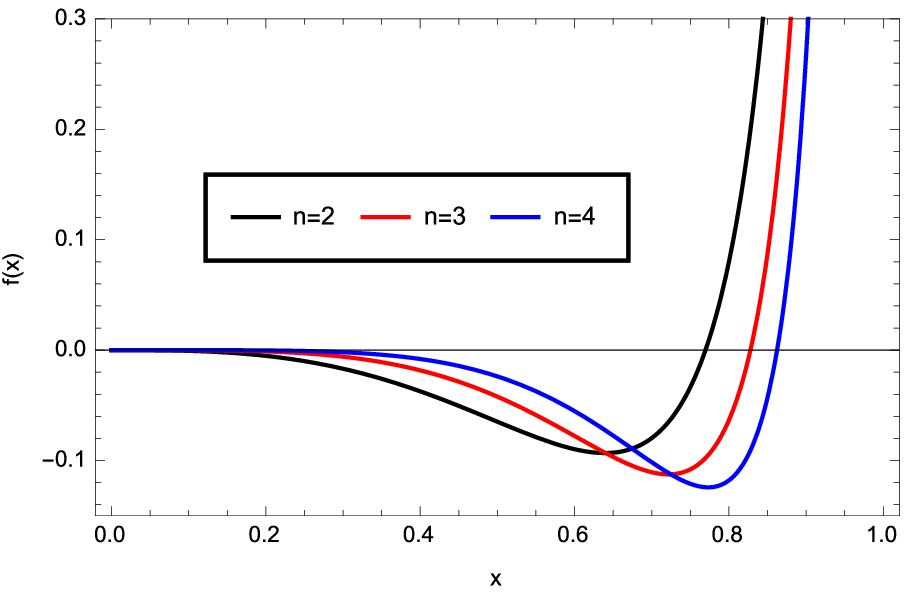}\hspace{0.5cm}
\includegraphics[width=7cm,keepaspectratio]{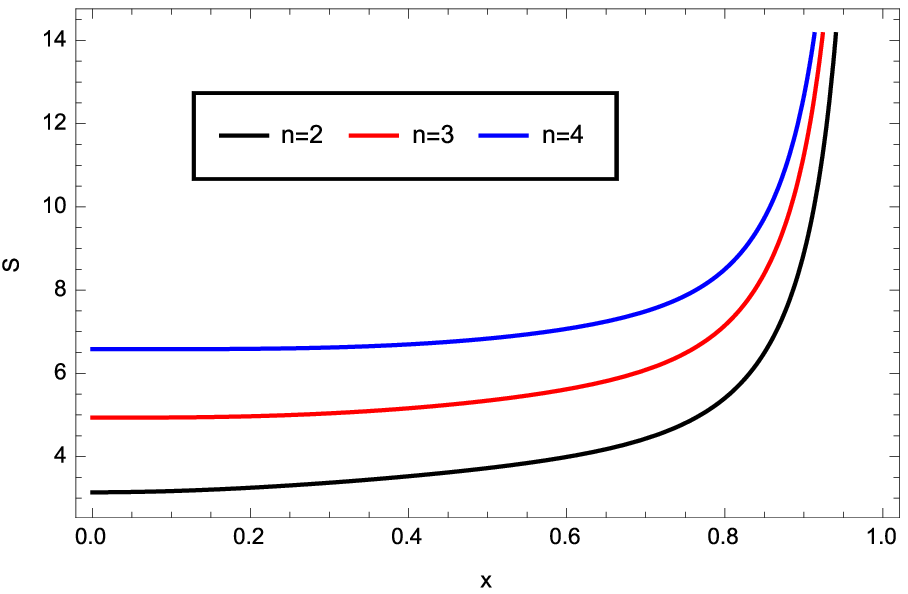}
\caption{ The left panel depicts $f(x)$ as functions of $x$ for $n+2$-dimensional RNdS black hole. The right panel depicts the whole entropy $S$ of the RNdS black hole in different dimensions. We have set $r_c=1$.}\label{figfx}
\end{figure}

\begin{figure}[!htbp]
\includegraphics[width=7cm,keepaspectratio]{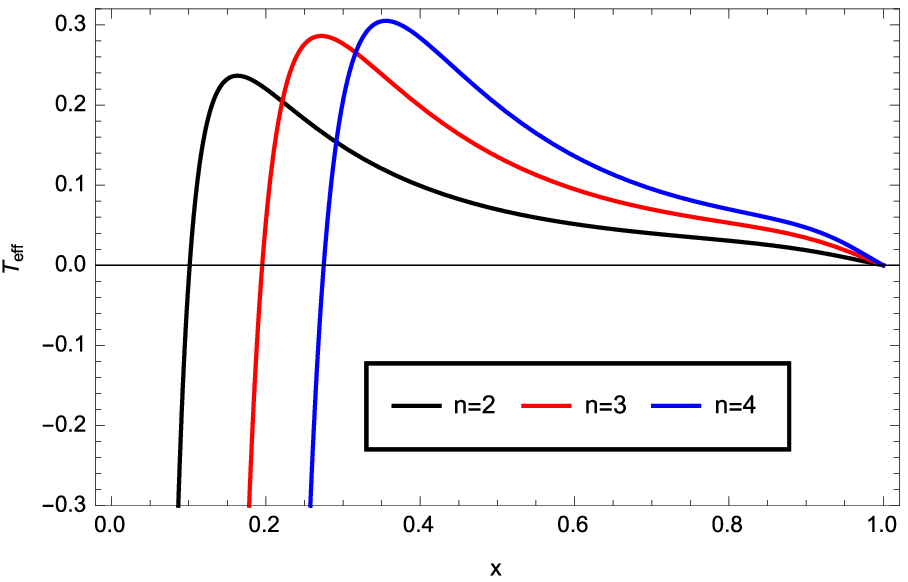}\hspace{0.5cm}
\includegraphics[width=7cm,keepaspectratio]{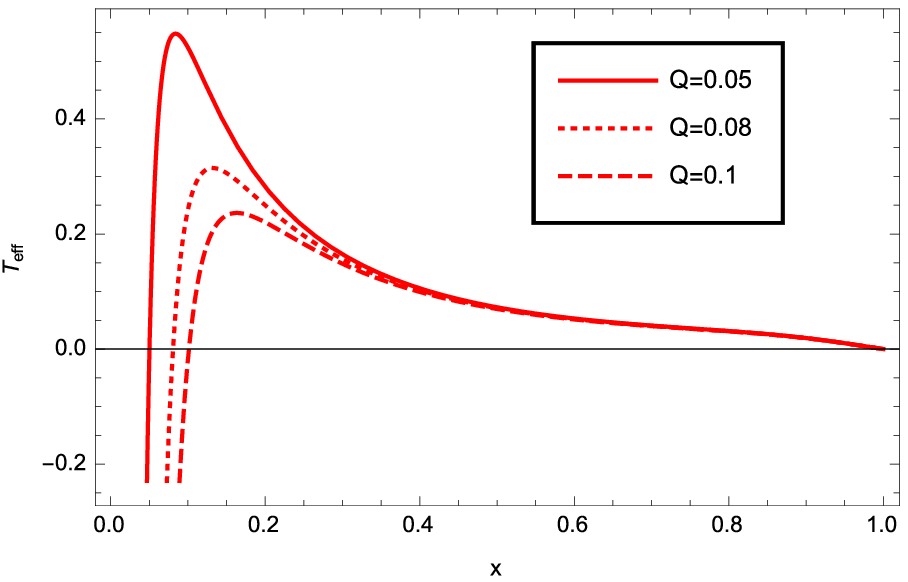}
\caption{The effective temperature as functions of $x$. The left panel depicts $T_{eff}$ at fixed $Q=0.05$. The right panel depicts depicts $T_{eff}$ at fixed $n=3$. We have set $r_c=1$.}\label{figTeff}
\end{figure}

As is shown in Fig.\ref{figfx}, the value of $f(x)$ does not vary monotonically. It first decreases as the $x$ increases, at some point it reaches a minimum and then begins to increase to the infinity at $x=1$.  The entropy $S$ increases with the spacetime deimension $n$ and diverges as $x \rightarrow 1$. We also depict the effective temperature $T_{eff}$ in Fig.\ref{figTeff}, from which we can see that $T_{eff} $ tends to zero as $x\to 1$,
namely the charged Nariai limit. Although this result does not agree with
that of Bousso and Hawking\cite{Bousso}, it is consistent with the entropy. Besides, the temperature has a maximum. The maximum of the temperature is dependent on the values of $n$ and $Q$. For larger $n$, the maximum lies at bigger $x$. And for larger $Q$, the maximum will be smaller.

\section{Phase transition in RN-dS black hole spacetime}

In analogy to the van der Waals liquid/gas system, one can analyze the black
hole thermodynamic system. One can derive the critical exponent,
Ehrenfest's equations. However, the de Sitter black hole cannot be
in thermodynamically equilibrium state in the usual sense due to the different temperatures on
the two horizons. From Eq.(3.6), the entropy of dS black holes should
contain an extra term $f(x)$. This result is obtained from the first law of
thermodynamics, which is the universal for usual thermodynamic system. Thus,
the entropy of the dS black hole we derived is closer to that of usual
thermodynamic system.

When $n=2$, the  effective potential is
\begin{equation}
\label{eq27}
\Phi _{eff} =\frac{Q(1-x^4)}{r_c x(1-x^3)},
\end{equation}
We can adjust the $T_{eff}$ as the function $\Phi _{eff}$, but not $Q$. So it is
\be
T_{eff}=\frac{\Phi ^2 \left(x^2+x+1\right)^2 \left[x^3 \left(x^4-3 x+3\right)-1\right]-x^2 (x+1)^2 \left(x^2+1\right)^2 \left[x^2 \left(x^3-3 x+3\right)-1\right]}{4 \pi  x^3 \left(x^3+x^2+x+1\right)^2 \left(x^4+1\right) r_c}.
\ee



\begin{figure}[!htbp]
\includegraphics[width=7cm,keepaspectratio]{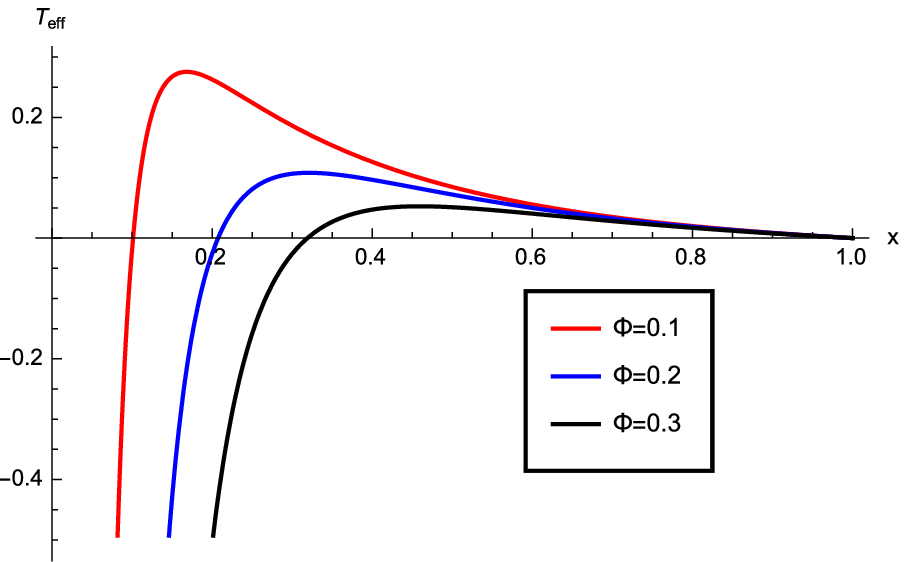}\hspace{0.5cm}
\includegraphics[width=7cm,keepaspectratio]{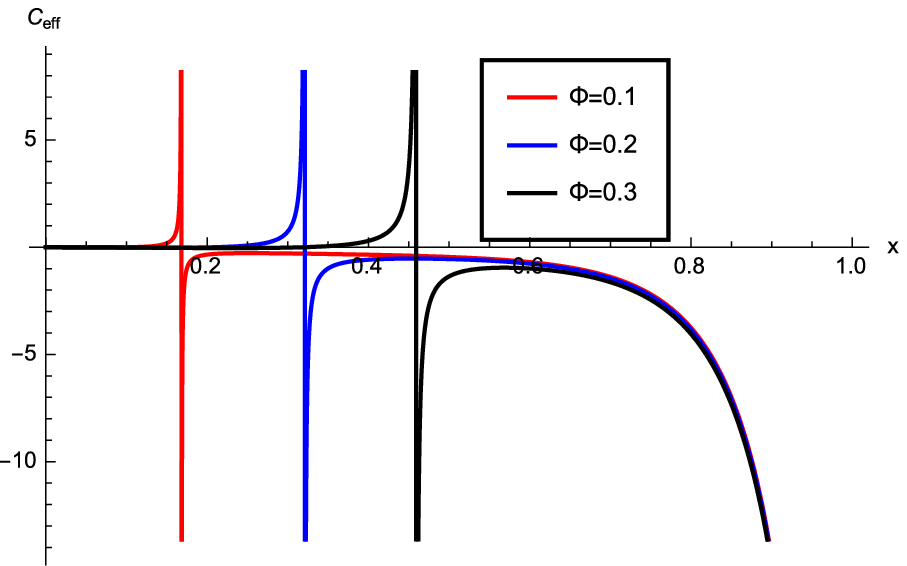}
\caption{The effective temperature and the effective heat capacity as functions of $x$ for different $\Phi_{eff}=0.1,~0.2,~0.3$ with fixed $n=2$. We have set $r_c=1$.}\label{figTph}
\end{figure}

The effective heat capacity can be defined as
\begin{equation}
\label{eq33}
C_{\Phi _{eff} } =T_{eff} \left( {\frac{\partial S}{\partial T_{eff} }}
\right)_{\Phi _{eff} }.
\end{equation}
In Fig.\ref{figTph}, we depict the effective temperature and the heat capacity at the fixed $\Phi_{eff}$ ensemble. It is shown that the heat capacity will diverge at the point where the effective temperature takes maximum. As the value of $\Phi_{eff}$ increases, the position of the divergent point moves right. Only on the left-hand side of that point, the heat capacity is positive. This means that the effective thermodynamic system is thermodynamically stable when the two horizons have a long way off.

The analog of volume expansion coefficient and analog of isothermal
compressibility are given by
\begin{equation}
\label{eq34}
\alpha =\frac{1}{Q}\left( {\frac{\partial Q}{\partial T_{eff} }}
\right)_{\Phi _{eff} } ,
\kappa _{T_{eff} } =-\frac{1}{Q}\left( {\frac{\partial Q}{\partial \Phi
_{eff} }} \right)_{T_{eff} } .
\end{equation}
They have the similar behaviors to that of the effective heat capacity.

We now exploit Ehrenfest's scheme in order to understand the nature of the
phase transition. Ehrenfest's scheme basically consists of a pair of
equations known as Ehrenfest's equations of first and second kind. For a
standard thermodynamic system these equations may be written as
\begin{equation}
\label{eq36}
-\left( {\frac{\partial \Phi _{eff} }{\partial T_{eff} }} \right)_S
=\frac{C_{\Phi _{eff} }^2 -C_{\Phi _{eff} }^1 }{T_{eff} V(\alpha ^2-\alpha
^1)}=\frac{\Delta C_{\Phi _{eff} } }{T_{eff} V\Delta \alpha },
\end{equation}
\begin{equation}
\label{eq37}
-\left( {\frac{\partial \Phi _{eff} }{\partial T_{eff} }} \right)_V
=\frac{\alpha ^2-\alpha ^1}{\kappa ^2-\kappa ^1}=\frac{\Delta \alpha
}{\Delta \kappa }.
\end{equation}
The subscript 1 and 2 represent phase 1 and 2 respectively. The new
variables $\alpha $ and $\kappa _{T_{eff} } $ correspond to the volume expansivity
and isothermal compressibility in statistical thermodynamics.

From the Maxwell's relations,
\begin{equation}
\label{eq38}
\left( {\frac{\partial Q}{\partial S}} \right)_{T_{eff} } =-\left(
{\frac{\partial T_{eff} }{\partial \Phi _{eff} }} \right)_Q ,
\quad
\left( {\frac{\partial S}{\partial Q}} \right)_{\Phi _{eff} } =-\left(
{\frac{\partial \Phi _{eff} }{\partial T_{eff} }} \right)_S
\end{equation}
substituting Eq.(\ref{eq38}) into Eqs.(\ref{eq36}) and (\ref{eq37}), we can obtain
\begin{equation}
\label{eq39}
\frac{\Delta C_{\Phi _{eff} } }{T_{eff} V\Delta \alpha }=-\left(
{\frac{\partial S}{\partial Q}} \right)_{\Phi _{eff} }^c ,
\quad
\frac{\Delta \alpha }{\Delta \kappa }=-\left( {\frac{\partial S}{\partial
Q}} \right)_{T_{eff} }^c .
\end{equation}
Note that the superscript ``c'' denotes the values of physical quantities at
a critical point in our article, while we find that
\begin{equation}
\label{eq40}
\left( {\frac{\partial S}{\partial Q}} \right)_{\Phi _{eff} } =\frac{\left(
{\frac{\partial S}{\partial x}} \right)_{r_c } \left( {\frac{\partial \Phi
_{eff} }{\partial r_c }} \right)_x -\left( {\frac{\partial \Phi _{eff}
}{\partial x}} \right)_{r_c } \left( {\frac{\partial S}{\partial r_c }}
\right)_x }{\left( {\frac{\partial Q}{\partial x}} \right)_{r_c } \left(
{\frac{\partial \Phi _{eff} }{\partial r_c }} \right)_x -\left(
{\frac{\partial \Phi _{eff} }{\partial x}} \right)_{r_c } \left(
{\frac{\partial Q}{\partial r_c }} \right)_x },
\end{equation}
\begin{equation}
\label{eq41}
\left( {\frac{\partial S}{\partial Q}} \right)_{T_{eff} } =\frac{\left(
{\frac{\partial S}{\partial x}} \right)_{r_c } \left( {\frac{\partial
T_{eff} }{\partial r_c }} \right)_x -\left( {\frac{\partial T_{eff}
}{\partial x}} \right)_{r_c } \left( {\frac{\partial S}{\partial r_c }}
\right)_x }{\left( {\frac{\partial Q}{\partial x}} \right)_{r_c } \left(
{\frac{\partial T_{eff} }{\partial r_c }} \right)_x -\left( {\frac{\partial
T_{eff} }{\partial x}} \right)_{r_c } \left( {\frac{\partial Q}{\partial r_c
}} \right)_x },
\end{equation}
When $\left( {\frac{\partial Q}{\partial x}}
\right)_{r_c } \left( {\frac{\partial T_{eff} }{\partial r_c }} \right)_x
-\left( {\frac{\partial T_{eff} }{\partial x}} \right)_{r_c } \left(
{\frac{\partial Q}{\partial r_c }} \right)_x \ne 0$, the critical points
satisfy
\begin{equation}
\label{eq42}
\left( {\frac{\partial T_{eff} }{\partial x}} \right)_{r_c } \left(
{\frac{\partial \Phi _{eff} }{\partial r_c }} \right)_x -\left(
{\frac{\partial \Phi _{eff} }{\partial x}} \right)_{r_c } \left(
{\frac{\partial T_{eff} }{\partial r_c }} \right)_x =0.
\end{equation}
Substituting Eq. (\ref{eq42}) into Eq. (\ref{eq40}), we have
\begin{equation}
\label{eq43}
\left( {\frac{\partial S}{\partial Q}} \right)_{\Phi _{eff} }^c =\left(
{\frac{\partial S}{\partial Q}} \right)_{T_{eff} }^c .
\end{equation}
So far, we have proved that both the Ehrenfest equations are correct at the
critical point. Utilizing Eq .(\ref{eq43}), the Prigogine--Defay (PD) ratio ($\Pi
)$can be calculated as
\begin{equation}
\label{eq44}
\Pi =\frac{\Delta C_{\Phi _{eff} } \Delta \kappa _{T_{eff} } }{T_{eff}^c
V^c(\Delta \alpha )^2}=1.
\end{equation}
Hence the phase transition occurring at $T_{eff} =T_{eff}^c $ is a second
order equilibrium transition . This is true in spite of the fact that the
phase transition curves are smeared and divergent near the critical point.

\section{Conclusions}

In this paper, we first propose the condition under which the black hole horizon and the cosmological horizon have the same temperature for the RN-dS black hole. We think that the entropy of these black holes with multiple horizons is not simply the sum of the entropies of every horizon, but should contains an extra contribution from the correlations between the horizons. On the basis of this consideration, we put forward the expression of the entropy. According to the effective first law of black hole thermodynamics, we can derive the effective temperature $T_{eff} $, the effective pressure $P_{eff}$ and the effective potential $\Phi _{eff}$. In the lukewarm case, the temperatures of the two horizons are the same. We conjecture that the effective temperature also takes the same value. According to this relation, we can obtain the differential equation for $f(x)$. Considering the reasonable boundary condition: $f(0)=0$, we can solve the differential equation exactly and obtain the $f(x)$.

In section IV, we analyzed the phase transition of the RN-dS black hole. Near the critical point, the heat capacity, the expansion coefficient and the isothermal
compressibility are all divergent, while at this point the entropy and the Gibbs free energy are continuous. Thus the phase transition at this point is of second order.  From Fig.\ref{figTph}, only when $x<x_c$ the heat capacity can be positive. This means that on the left-hand side of the divergent point, the effective thermodynamic system is locally thermodynamically stable.

We anticipate that study on the thermodynamic properties of the black holes in de Sitter space can shed light on the classical and quantum properties of de Sitter space.

\section{acknowledgments}
This work is supported in part by the National Natural Science Foundation
of China (Grant No.11475108).

\end{document}